\documentclass[aps,11pt,onecolumn,preprintnumbers,amsmath,amssymb]{revtex4}

\usepackage[english]{babel}
\usepackage{graphicx}
\usepackage{epstopdf}
\usepackage{color}
\usepackage{ragged2e}
\usepackage{float}

\begin{document}

\newcommand{\unit}[1]{\:\mathrm{#1}}            
\newcommand{\To}{\mathrm{T_0}}
\newcommand{\Tp}{\mathrm{T_+}}
\newcommand{\Tm}{\mathrm{T_-}}
\newcommand{\EST}{E_{\mathrm{ST}}}
\newcommand{\Rp}{\mathrm{R_{+}}}
\newcommand{\Rm}{\mathrm{R_{-}}}
\newcommand{\Rpp}{\mathrm{R_{++}}}
\newcommand{\Rmm}{\mathrm{R_{--}}}
\newcommand{\ddensity}[2]{\rho_{#1\,#2,#1\,#2}} 
\newcommand{\ket}[1]{\left| #1 \right>} 
\newcommand{\bra}[1]{\left< #1 \right|} 

\title{Dipolar interactions between field-tuneable, localized emitters in van der Waals heterostructures}
\author{Weijie Li$^{\dagger}$}
\author{Xin Lu$^{\dagger}$}
\author{Sudipta Dubey}
\author{Luka Devenica}
\author{Ajit Srivastava$^{*}$}\affiliation{Department of Physics, Emory University, Atlanta 30322, Georgia, USA}
\maketitle
\justify
$^{\dagger}$These authors contributed equally to this work.\\
$^{*}$Correspondence to: ajit.srivastava@emory.edu




\textbf{While photons in free space barely interact, matter can mediate interactions between them resulting in optical nonlinearities. Such interactions at the single-quantum level result in an on-site photon repulsion~\cite{BirnbaumNature2005, FaraonNPhys2008}, crucial for photon-based quantum information processing and for realizing strongly interacting many-body states of light~\cite{ImamogluPRL1997,ChangNP2014, CarusottoRMP2013, ChangNPhys2008, LukinPRL2001}. Here, we report repulsive dipole-dipole interactions between electric field tuneable, localized interlayer excitons in MoSe$_2$/WSe$_2$ heterobilayer. The presence of a single, localized exciton with an out-of-plane, non-oscillating dipole moment increases the energy of the second excitation by $\sim$ 2 meV -- an order of magnitude larger than the emission linewidth and corresponding to an inter-dipole distance of $\sim$ 5 nm. At higher excitation power, multi-exciton complexes appear at systematically higher energies. The magnetic field dependence of the emission polarization is consistent with spin-valley singlet nature of the dipolar molecular state. Our finding is an important step towards the creation of excitonic few- and many-body states such as dipolar crystals with spin-valley spinor in van der Waals (vdW) heterostructures.}

Optical response in atomically thin layered semiconductors is determined by excitons and other excitonic complexes such as trions and biexcitons which are strongly bound due to increased Coulomb interactions in truly 2D limit~\cite{ChernikovPRL2014, HePRL2014}. In addition, due to the type-II band alignment in heterobilayer of MoSe$_2$/WSe$_2$, an interlayer exciton comprising of an electron in the MoSe$_2$ layer and hole in the WSe$_2$ layer is found to be stable and long-lived~\cite{RiveraNComm2015, TranNature2019,RiveraScience2016,FoglerNComm2014}. As shown in Fig.~1a, due to the spatial separation of electron and hole, the interlayer exciton carries a static, out-of-plane electric dipole moment which allows for the tuning of its energy by an external electric field ($E$). The orientation of this dipole is fixed by the ordering of MoSe$_2$ and WSe$_2$ layers and hence leads to a repulsive interaction between interlayer excitons. 

\begin{figure}
\includegraphics[scale=0.8]{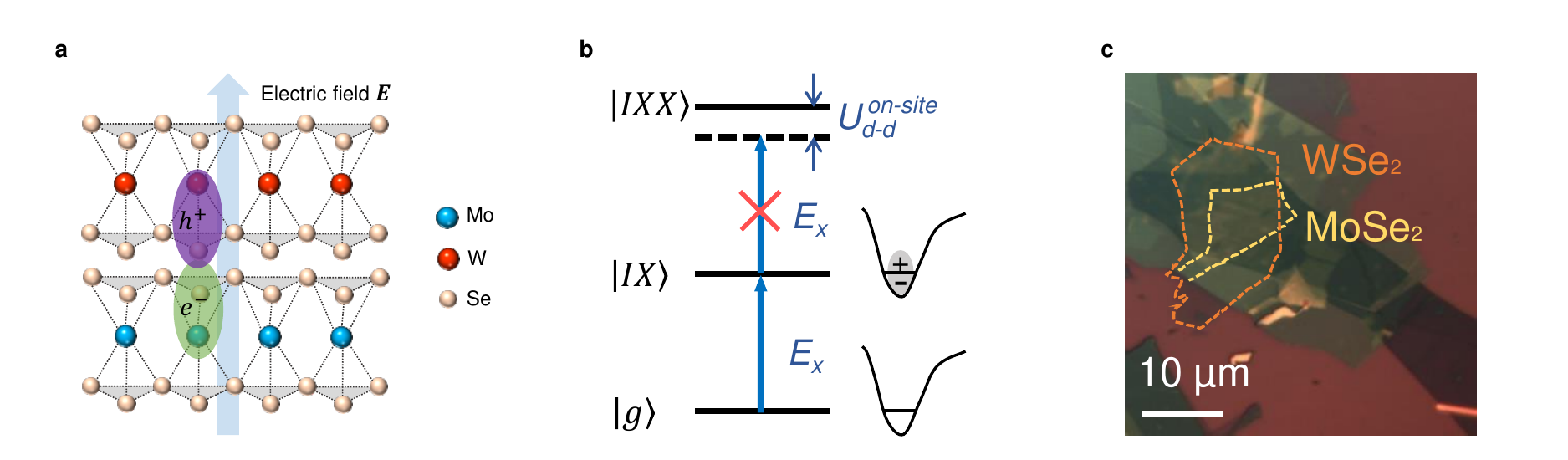}
\caption{{\bf Interlayer exciton dipoles in WSe$_2$/MoSe$_2$ heterostructure.} {\bf a,} A schematic showing the interlayer exciton in WSe$_2$-MoSe$_2$ heterobilayer under an external electric field $E$. Due to the type-II band alignment, electron and hole are separated in MoSe$_2$ and WSe$_2$, respectively, forming a permanent out-of-plane dipole. The dipole energy red-shifts (blue-shifts) when $E$ is parallel (anti-parallel) to the direction of dipole. {\bf b,}  Energy diagram of localized interlayer exciton and biexciton in a potential well. The energy of biexciton is raised up by on-site dipole-dipole interaction $U_{\mathrm{dd}}^{\mathrm{on-site}}$. {\bf c,} An optical image of WSe$_2$/MoSe$_2$ heterobilayer with graphite bottom gate. Monolayer WSe$_2$ (MoSe$_2$) is depicted in orange (yellow) dashed line. The final device has graphite bottom and top gates with h-BN as dielectric on both sides.}
\end{figure}

This dipolar interaction is potentially interesting for inducing effective photon-photon interaction and making the heterobilayer a nonlinear optical medium for efficient photon switching applications. Furthermore, repulsive dipolar interactions could also lead to exotic many-body states such as self-assembled dipolar crystals with spin-valley degree of freedom at a critical density of excitons~\cite{RablPRA2007, LahayeRPP2009}. As a signature of nonlinearity, dipolar repulsion should result in an exciton density dependent blueshift of the emission energy with increasing incident power. While such a behavior has been previously reported~\cite{CiarrocchiNP2019,Philipp2DMaterials2017}, a discernible shift of the emission energy, comparable to the linewidth, occurs only for a large number of excitons ($\sim$ 10$^4$-10$^5$) and at the highest incident power. 

In order to see a quantum nonlinearity where the presence of merely one additional exciton drastically modifies the optical response, the effective dipole-dipole interaction (U$_{\mathrm{dd}}$) must be larger than the linewidth~\cite{ChangNP2014}. As the expected blueshift is U$_{\mathrm{dd}}$ $\sim$ $1/r_{\mathrm{ex}}^3$ where $r_{\mathrm{ex}}$ is the interexcitonic distance, localized interlayer excitons are a good candidate to observe this effect. Localized excitons in monolayer WSe$_2$ have been shown to be single photon emitters with sharp linewidths~\cite{AtatureNRM2018, AharonovichNP2016} and can host a single charge and spin~\cite{LuNNano2018, BrotonsNNano2019}. Very recently, localized interlayer excitons with sharp linewidths were reported in vdW heterostructures~\cite{SeylerNature2019}. As described in Fig.~1b, there is an on-site energy cost ($U_{\mathrm{dd}}^{\mathrm{on}-\mathrm{site}}$) for creating two interlayer excitons within the same trap ($|IXX\rangle$) compared to a single excitation ($|IX\rangle$) with energy $E_X$. The situation is then reminiscent of dipole blockade in Rydberg atoms~\cite{SaffmanRMP2010} albeit with a much smaller dipole moment and hence requiring tighter localization. In addition to increasing $U_{\mathrm{dd}}$, localized interlayer excitons should also serve as quantum emitters consisting of a single dipole with tuneable emission energy in a perpendicular electric field.

To demonstrate dipole-dipole interactions in localized interlayer excitons, we fabricated a MoSe$_2$/WSe$_2$ heterobilayer encapsulated in hexagonal-boron nitride layers with graphite top and bottom gates, as shown in Fig.~1c (see Methods). Fig.~2a shows the low temperature ($\sim$ 4~K) photoluminescence (PL) spectra of our sample with emission from interlayer exciton clearly present at lower energy ($\sim$ 1.35 eV) compared to intralayer exciton peaks of MoSe$_2$ (WSe$_2$) at 1.65 eV (1.7 eV), consistent with previous studies~\cite{RiveraNNano2018}. Reflectance spectra of the heterobilayer region shown in Fig.~2b exhibits a redshift of intralayer exciton resonances compared to the monolayer regions as is expected from the interaction between the two monolayers~\cite{SchaibleyNComm2016}. The photoluminescence excitation (PLE) spectroscopy of the lower energy PL peaks shows resonances at excitation energies corresponding to MoSe$_2$ and WSe$_2$ excitons, further confirming the interlayer nature of the redshifted peaks (Fig.~2c).

\begin{figure}
\includegraphics[scale=0.2]{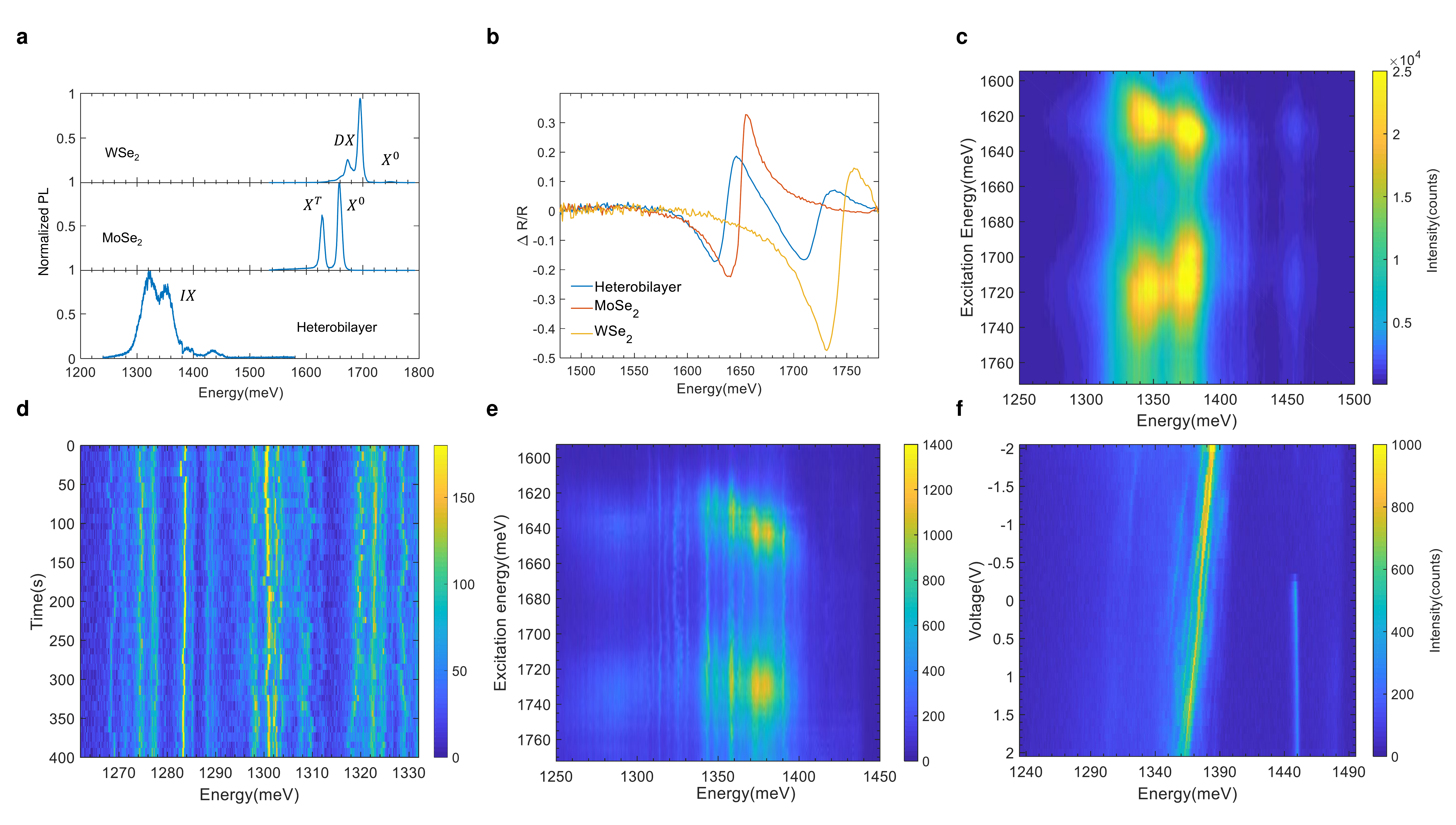}
\caption{{\bf Excitons from WSe$_2$/MoSe$_2$ heterostructure at low temperature ($\sim$4~K).}   {\bf a,} Normalized photoluminescence (PL) spectra. Top: Emission from defect band (DX) dominates in WSe$_2$. Exciton $X^0$ is observed at $\sim$1750~meV. Middle: Monolayer MoSe$_2$ shows two prominent peaks at 1658~meV and 1628~meV, corresponding to neutral exciton $X^0$ and trion $X^T$, respectively. Bottom: Emission from interlayer excitons appears at lower energy (1250 - 1450~meV). {\bf b,}  Reflectance contrast $\Delta R/R$ spectra. $\Delta R/R$ = ($R_{\mathrm{sample}}$ -  $R_{\mathrm{substrate}}$ ) /$R_{\mathrm{substrate}}$. The exciton resonances in monolayers are consistent with the observation from PL spectroscopy in panel {\bf a} . Compared to monolayers, the intralayer excitons in heterobilayer  red-shift and broaden.  {\bf c,} Photoluminescence excitation (PLE) spectroscopy of interlayer excitons. {\bf d,} Time-trace PL emission of localized interlayer excitons. {\bf e,} PLE of localized emitters. All the quantum emitters show local resonances around $\sim$1640~meV (MoSe$_2$ resonance) and $\sim$~1730~meV (WSe$_2$ resonance), consistent with the nature of interlayer excitons.  {\bf f,} Electric field tuning of a localized interlayer exciton.  As bottom gate voltage sweeps from -2 to +2 V, the localized interlayer exciton at  $\sim$1380~meV  exhibits a red-shift of $\sim$20~meV, confirming the existence of an out-of-plane dipole. As a comparison, the localized intralayer exciton  at  $\sim$1450~meV does not show electric field tunability. Excitation laser is linearly-polarized. Wavelength $\lambda$ = 633~nm in panel {\bf a}, $\lambda$ = 770~nm in panel {\bf d}, $\lambda$ = 758~nm in panel {\bf f}. Incident power $P$ =  2~$\mu$W, 5~$\mu$W and 20~$\mu$W from top to bottom in panel {\bf a},  $P$ = 10~nW in panel {\bf d}, $P$ = 210~nW in panel {\bf f}.}
\end{figure}

In order to observe localized interlayer excitons, we switch to quasi-resonant excitation close to MoSe$_2$/WSe$_2$ exciton and use low excitation power $\sim$ nW (Methods). As Fig.~2d shows, we observe sharp, spatially localized peaks with linewidths as low as 110 $\mu$eV in the energy range of 1260 to 1330~meV (see Supplementary). Furthermore, these sharp emission peaks show spectral jittering which is characteristic of localized quantum emitters. Since the energy of the sharp peaks lies in the range of interlayer PL at higher power and because intralayer WSe$_2$ and MoSe$_2$ localized excitons typically exhibit energy higher than 1560~meV~\cite{BrannyAPL2016}, we believe that such low-energy peaks should be related to interlayer exciton. To further confirm our claim, we perform PLE spectroscopy and find resonances at both MoSe$_2$ ($\sim$1640~meV) and WSe$_2$ ($\sim$~1730~meV) excitons (Fig.~2e), indicating the interlayer nature of these peaks.

\begin{figure}
\includegraphics[scale=0.85]{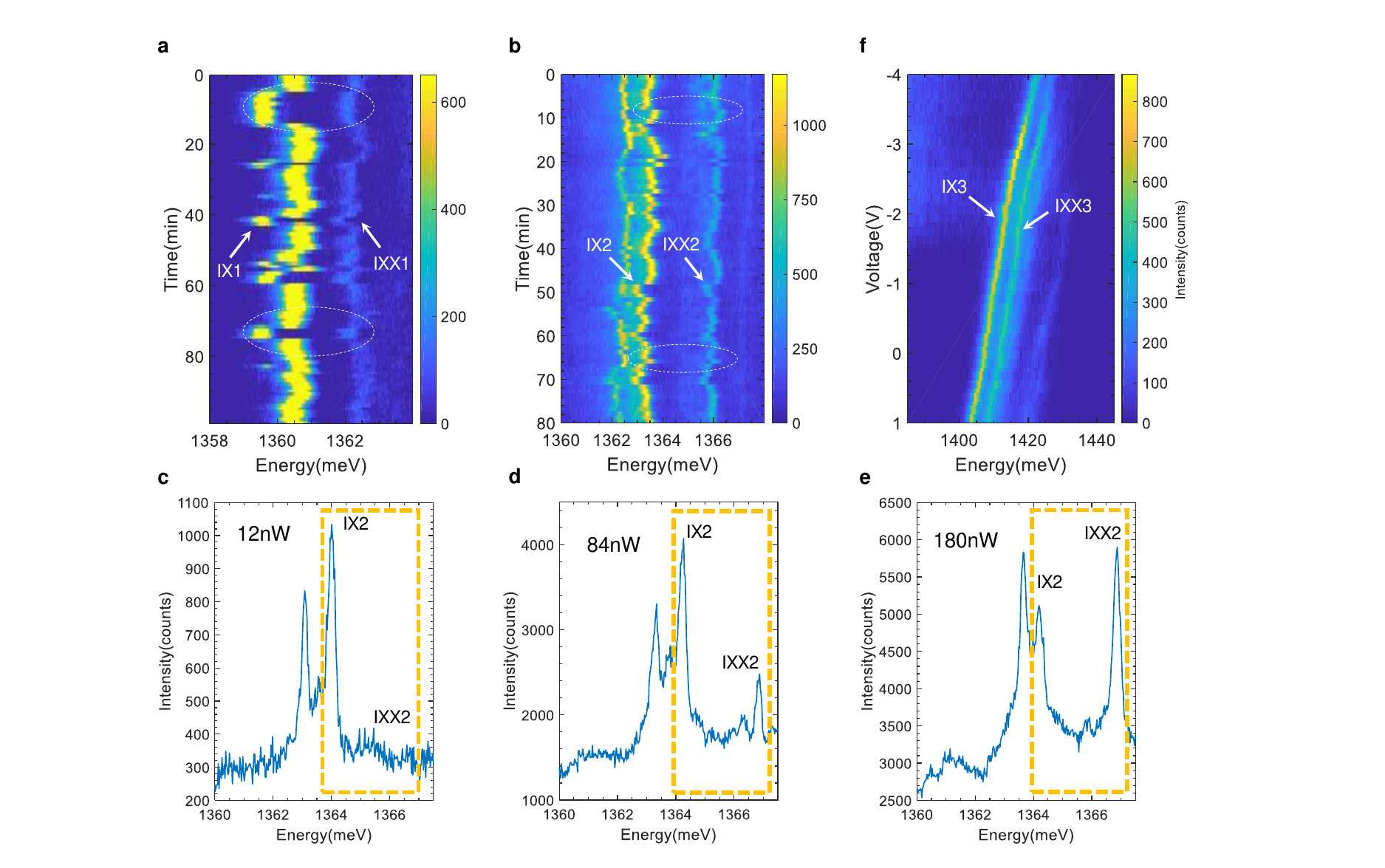}
\caption{{\bf Signature of localized interlayer biexcitons.} {\bf a, b,} Time-trace of PL emission of interlayer excitons and biexcitons. White arrows and dotted circles highlight the same spectral jittering patterns from IX1-IXX1 (a) and IX2-IXX2 (b), indicating that they originate from the same quantum emitter. {\bf c-e} PL spectra of IX2-IXX2 under different excitation powers, 12~nW (c), 84~nW (d) and 180~nW (e). Only exciton IX2 shows up at low power (12~nW). Biexciton IXX2 appears at intermediate power (84~nW) and dominates at high power (180~nW). {\bf f,} Electrical tuning of interlayer exciton IX3 and biexciton IXX3. The two peaks exhibit the same tuning rate of $\sim$310~meV nm V$^{-1}$. Excitation laser is linearly-polarized. Wavelength $\lambda$ = 745~nm in panel {\bf a-d} , $\lambda$ = 758~nm in panel {\bf f}. Incident power $P$ = 20~nW in panel {\bf a}, $P$ = 250~nW in panel {\bf b}, $P$ = 4~$\mu$W in panel {\bf f}.}
\end{figure}

With electron and hole located in different layers due to the type-II band alignment, interlayer exciton has an out-of-plane permanent dipole and its energy should blueshift (redshift) by an amount $\Delta U = -p\cdot \Delta E$ when the electric field ($E$) is anti-parallel (parallel) to the the direction of dipole $p$. The dipole moment in our sample is 0.7~nm $\cdot$ $e$ and points from MoSe$_2$ (bottom layer) to WSe$_2$ (top layer). As the bottom gate voltage changes from -2 to +2~V, $\Delta E$ (pointing up) =  +4~eV $\cdot$ 80~nm (the estimated thickness between top and bottom gates). We thus expect the energy of interlayer exciton to decrease by $\sim$22 meV. The observed red-shift of $\sim$20~meV (tuning rate of 400~meV nm V$^
{-1}$) in Fig.~2f is in consistent with the analysis ~\cite{CiarrocchiNP2019}. Meanwhile, a localized peak at 1450~meV, possibly from MoSe$_2$, hardly shifts with $E$. This is in agreement with the absence of out-of-plane dipole moment for an intralayer exciton. The $E$-tunability of sharp peaks unambiguously demonstrates that they arise from localized interlayer excitons. It is very probable that, like their intralayer counterparts, interlayer excitons also get trapped in shallow potentials due to strain or defect potential on a length scale larger than interlayer exciton Bohr radius ($\sim$ 2 nm)~\cite{AtatureNRM2018}. A desirable property of these quantum emitters is that their energy can be tuned by more than 100 times their linewidth. Thus, we can conclude that localized interlayer excitons are excellent candidates to study dipolar interactions in 2D layered materials.

Having established that we have observed localized interlayer exciton with an out-of-plane dipole, we investigate their dipole-dipole interactions. To this end, we slightly increase the excitation laser power for larger density of excitons. Fig.~3a is the time-trace PL emission of two peaks at $\sim$1360 and $\sim$1362~meV which show the same spectral jittering pattern, as highlighted by white arrows and dotted circles. This behavior suggests that the two peaks belong to the same localized interlayer exciton. Similar feature is also observed in other localized interlayer excitons, with energy spacing between the two peaks varying from 1 to 5~meV (Fig.~3b and supplementary). We notice that such synchronized spectral jittering is not shown by all the peaks in our collection spot-size. For example, the lowest-energy peak in Fig.~3b exhibits a different pattern. In localized intralayer excitons, a doublet peak structure showing similar synchronized jittering is seen and arises from electron-hole (e-h) exchange interaction which causes a fine structure splitting~\cite{AtatureNRM2018}. However, we can rule out the possibility of such fine structure splitting, as the e-h exchange interaction is strongly quenched in interlayer exciton due to the separation of carriers into distinct layers. Thus, exciton complexes such as charged exciton and biexciton could be the possible origin of the two peak structure.

Excitation power-dependence of emission intensity is an ideal technique to distinguish between charged exciton and biexciton. Fig. 3c-e shows PL spectra at different incident powers. At the lowest power, we only observe the red peak. As the power is increased, the blue peak starts to show up at intermediate power and becomes stronger than the red peak at higher power. This strongly suggests that the blue peak is possibly a biexciton. We thus assign the two peaks as IX and IXX, respectively. We further plot the integrated intensity of each emission peak as a function of excitation power, and fit the data with a power law function, $I$ $\propto$ $P^\alpha$  (Supplementary). The red peak IX2  exhibits a power-law behavior with $\alpha$ $\sim$ 1.0, and the blue peak IXX2 shows a super-linear power dependence with $I$ $\propto$ $P^{2\alpha}$. The super-linear power dependence is consistent with our assignment that the blue peak is a biexciton although the confidence in the value of $\alpha$ is poor due to a limited range of powers where the peaks can observed prior to saturation of their emission intensities. Fig.~3f shows that interlayer exciton and biexciton exhibiting the same $E$-tuning rate of $\sim$310~meV nm V$^{-1}$ suggesting that they both carry a dipole moment.

While the biexciton in monolayer WSe$_2$ emits at lower energy from PL spectroscopy because of a finite, positive binding energy~\cite{HeNComm2016}, the energy of interlayer biexciton state is raised up by $U_{\mathrm{dd}}$ due to dipolar repulsion (Fig.~4a). Emission from biexciton is thus blue-shifted with respect to exciton (Fig.~3 and Supplementary). The different energy spacing (1-5~meV) between exciton and biexciton indicates that the dipolar interaction varies among different localized interlayer excitons. As the dipole moment can be assumed to be constant given by the separation of 0.7 nm between the two monolayers, variation in $U_{\mathrm{dd}}$ must arise from difference in confinement lengths and consequently interexcitonic distances. To estimate the confinement length from $U_{\mathrm{dd}}$, we assume that the interlayer excitons are confined in a harmonic trap with a width larger than the excitonic Bohr radius such that the dipoles can be treated as point particles without considering their internal structure. As the trap is loaded with an additional exciton, the center-of-mass (COM) wavefunction of each exciton is squeezed to avoid overlap and lower the dipolar repulsion. The modified COM wavefunction of each exciton is no longer that of the ground state but has weight from higher energy excited states. This results in the increase of kinetic energy of the two-particle system. The interexcitonic distance can then be calculated by minimizing the total energy which includes $U_{\mathrm{dd}} = p^2 / (\epsilon_r r_{\mathrm{ex}}^3)$. For an energy difference of 2 meV between the exciton ($|IX\rangle \rightarrow |0\rangle$) and biexciton ($|IXX\rangle \rightarrow |IX\rangle$) emission peaks, we obtain a confinement length of $\sim$ 5 nm which is larger than the Bohr radius, validating our assumption (see Supplementary). We remark that one of the peak from another group shows power dependence with $I$ $\propto$ $P^{3\alpha}$, and possibly corresponds a triexciton. Indeed, at higher incident power, new peaks appear at even higher energy compared to the biexciton peak. We assignthem to multi-exciton complexes with a regular arrangement of excitons resembling dipolar lattice, which reduces dipolar repulsion~\cite{SchinnerPRL2013} (see Supplementary).

In addition to $U_{\mathrm{dd}}$, depending on the species of biexciton i.e., same valley ($X_+X_+$ or $X_-X_-$) or opposite valley ($X_+X_-$) interlayer excitons, exchange interaction, $U_{\mathrm{ex}}$, should affect the energy of biexciton as well. Here, $X_\pm$ denotes the exciton with electron and hole in the $\pm$K valley. The overall wavefunction of the two-exciton state is antisymmetric in the spatial coordinates to reduce the dipolar repulsion. The bosonic nature of the exciton then implies that the singlet arrangement of spin-valley or the opposite valley biexciton $X_+X_-$ has lower energy whereas the same valley excitons $X_\pm X_\pm$ have energy further increased by $U_{\mathrm{ex}}$. 

\begin{figure}
\includegraphics[scale=0.8]{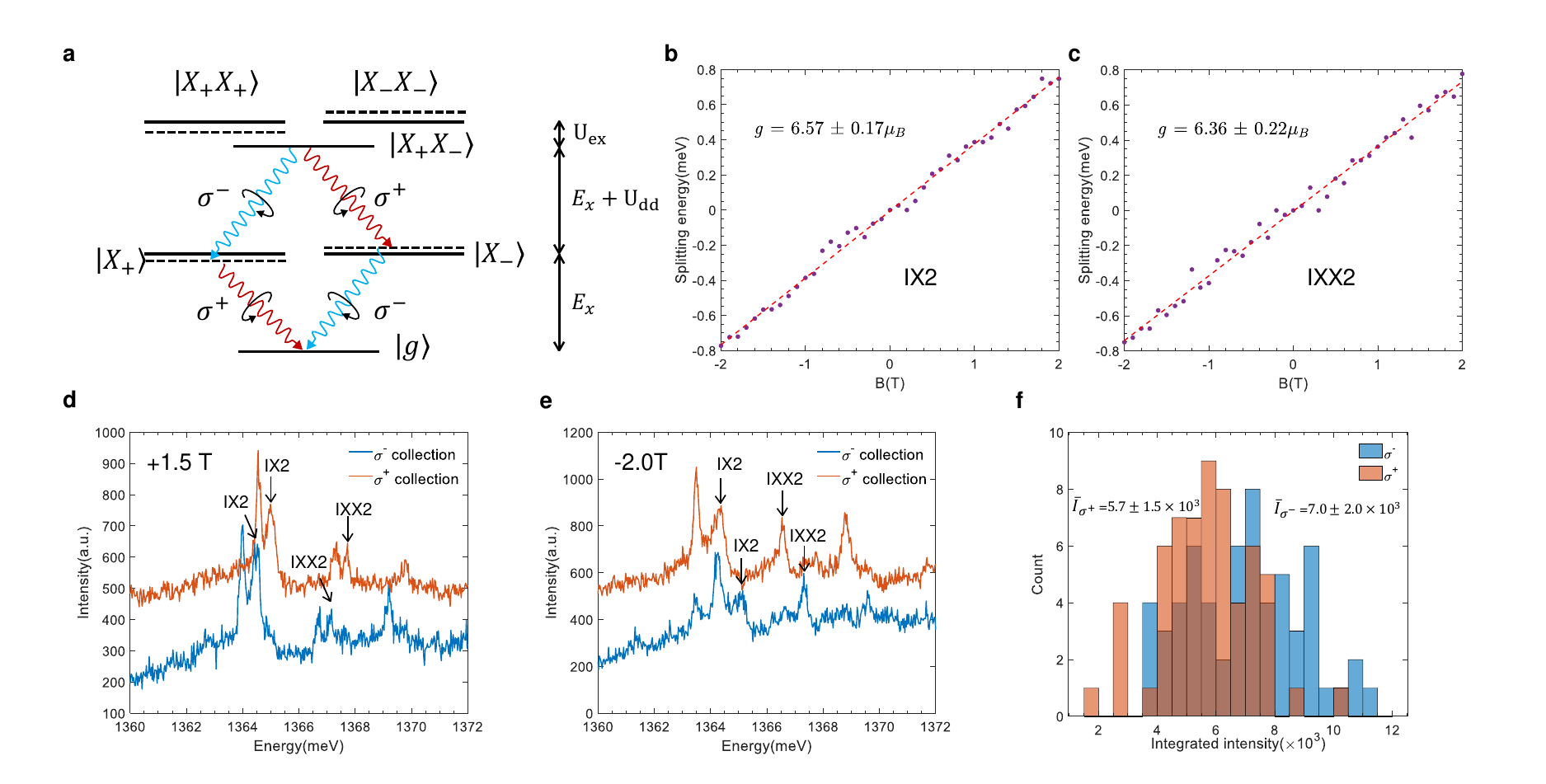}
\caption{{\bf Magnetic field dependence of localized interlayer excitons and biexcitons.} {\bf a,} Optical selection rule of IX and IXX under magnetic field. $|X_{+}\rangle$ and $|X_{-}\rangle$ states, as well as  $|X_{+}X_{+}\rangle$ and $|X_{-}X_{-}\rangle$, are doubly-degenerate at zero magnetic field, $B$ (solid lines). Lifted degeneracy of $|X_{+}\rangle$ and $|X_{-}\rangle$ under finite $B$ (dashed  lines) leads to the splitting of $|X_{+}X_{-}\rangle$ emission. Red (Blue) peaks of exciton and biexciton are both $\sigma^+$  ($\sigma^-$) polarized.  Degeneracy of $|X_{+}X_{+}\rangle$ and $|X_{-}X_{-}\rangle$ is also broken under $B$. {\bf b,c,} $B$-dependent splitting energies of IX2 (b) and IXX2 (c). Exciton and biexciton exhibit the same value of $g$-factor, $\sim$6.5. {\bf d,e,} Polarization-resolved PL spectra of IX2 and IXX2 at +1.5~T(d) and -2.0~T(e).  Polarization of the biexciton peaks follows that of the exciton. {\bf f,} Histogram of biexction intensity under $\sigma^+$  and $\sigma^-$ excitation configurations at -0.5~T. Biexciton is stronger under $\sigma^-$ excitation, consistent with the fact that the DCP of exciton is lower when being excited with $\sigma^-$ polarized light (see Supplementary Information). Excitation wavelength $\lambda$ = 745~nm in panel {\bf b-f}. Incident power $P$ = 250~nW (linearly-polarized) in panel {\bf b-e}, $P$ = 150~nW (circularly-polarized) in panel {\bf f}.}
\end{figure}

Much like the intralayer excitons, the interlayer excitons $X_\pm$ couple to circularly-polarized light with opposite helicity ($\sigma^\pm$) following the optical selection rule~ \cite{YuSciAdv2017, WuPRB2018}.  As we only observe one extra peak appearing with larger power, we tentatively suppose this peak is the lower energy $X_+X_-$ biexciton rather than the degenerate $X_+X_+$ and $X_-X_-$. $U_{\mathrm{ex}}$ needs to be overcome for the observation of $X_+X_+$ and $X_-X_-$ while the emission in PL typically arises only from the lowest energy state.  In other words, if the same valley biexciton is observed in PL emission, one expects to observe $X_+X_-$ simultaneously at the red side of the degenerate $X_+X_+$ and $X_-X_-$ states which does not seem to be case. The degeneracy of $X_+X_+$ and $X_-X_-$, as well as $X_+$ and $X_-$, is lifted under finite magnetic field, $B$ (dashed line in Fig.~4a). As illustrated in Fig.~4a, the degeneracy lifting of exciton states causes the PL emission from $X_+X_-$ split into two peaks, though $X_+X_-$ state is hardly affected under $B$. Co-polarized emission is expected with all red (blue) peaks emitting $\sigma^+$ ($\sigma^-$) polarized light, as shown in the schematic of Fig.~4a. The situation in Fig.~4a should be contrasted with that of intralayer biexciton where a finite e-h exchange splits the degeneracy of $X_\pm X_\mp$ states to make them linearly polarized. As a result, the biexciton cascade of the intralayer exciton yields maximally polarization entangled pairs of photons but the time averaged fidelity of the entanglement is reduced due the e-h splitting~\cite{HeNComm2016,HuberPRL2018}. Due to the absence of e-h exchange in interlayer excitons, localized interlayer biexcitons are ideal sources of maximally entangled photons. While the emission rate of our localized interlayer excitons (20 counts/s, see Methods) is slightly larger than the previously reported values~\cite{SeylerNature2019}, it is still too weak to detect entangled photons in a photon coincidence measurement. Recently, layer-hybridized interlayer excitons were reported in vdW heterobilayers where one of the charge carriers is delocalized in both layers thereby  increasing the oscillator strength ~\cite{AlexeevNature2019, HsuARXIV2019}.

Next, we proceed to analyze IXs and IXXs under $B$. Fig.~4b $\&$ 4c shows that the IX2 and IXX2 have the same $g$-factor of $\sim$6.5 (also see Supplementary for another quantum emitter group), which implies that the addition of second exciton has negligible impact on the wavefunction corresponding to the relative motion of electrons and holes. This is not surprising because the interlayer exciton Bohr radius ($\sim$ 2 nm) is estimated to be the smaller length scale compared to confinement length of the trap ($\sim$ 5 nm). As a result, the two excitons in the trap would avoid spatial overlap to reduce $U_{\mathrm{dd}}$ without changing the relative motion of the constituent electron-hole pair which determines the $g$-factor. From polarization-resolved PL measurements (Fig.~4d $\&$ 4e), we observe that both red and blue peaks of IXX2 are co-polarized with those of IX2, consistent with the energy diagram in Fig.~4a. 
To further confirm that the biexciton peak corresponds to a spin-valley singlet configuration as in $X_+X_-$/$X_-X_+$, we analyze the intensity of emission of the two circularly polarized components. A polarization resolved coincidence measurement of the red and blue photons in the biexciton cascade emission unequivocally determines the spin-valley configuration biexciton state. However, this measurement is made difficult due to the weak emission and unsuitable wavelength range for detection with silicon avalanche photodiodes. 

In principle, the intensity of $X_+X_-$ ($I_{X_+X_-}$) should depend on the product $I_{X_+}$ $I_{X_-}$, and $I_{X_+X_+}$ ($I_{X_-X_-}$) would be higher if $I_{X_+}$ $I_{X_+}$ ($I_{X_-}$ $I_{X_-}$) is larger. We use excitation polarization to control the degree of circular polarization (DCP) of exciton, such that the relative intensity between $X_+$ and $X_-$ is modulated while keeping the total intensity unchanged.  DCP is obtained by calculating ($I_{\sigma^+}$ - $I_{\sigma^-}$) / ($I_{\sigma^+}$ + $I_{\sigma^-}$), where a positive (negative) value means the peak is $\sigma^+$ ($\sigma^-$) polarized. DCP of exciton IX2 under $\sigma^-$ and $\sigma^+$ excitations is -0.19 and -0.29 at $B$ = -0.5~T, respectively. Because the exciton is less circularly-polarized with $\sigma^-$ pumping, we expect IXX2 to be stronger if it is $X_+X_-$. Histogram based on more than 50 polarization-resolved spectra for each polarization confirms our analysis. Fig.~4f shows that the mean integrated intensity from $\sigma^-$ excitation is $\sim$7000, and that from $\sigma^+$ excitation is only $\sim$5700. To further support our assignment, we calculate the ratio of $I_{XX}$ $/$ $I_{X_+}$ $I_{X_-}$, $I_{XX}$ $/$ $I_{X_+}$ $I_{X_+}$  and $I_{XX}$ $/$ $I_{X_-}$ $I_{X_-}$ (Supplementary Table I). For $X_+X_-$ biexciton, $I_{XX}$ should be proportional to $I_{X_+}$ $I_{X_-}$, regardless of the exciton polarization. However, such a proportionality does not hold for $I_{XX}$ $/$ $I_{X_+}$ $I_{X_+}$  and $I_{XX}$ $/$ $I_{X_-}$ $I_{X_-}$, confirming our assumption. We characterize the difference of ratio (R) between $\sigma^+$ and $\sigma^-$ excitations by calculating 2($R$($\sigma^+$) - $R$($\sigma^-$)) / ($R$($\sigma^+$) + $R$($\sigma^-$)). The difference of ratio is one order small for $I_{XX}$ $/$ $I_{X_+}$ $I_{X_-}$ compared to that of $I_{XX}$ $/$ $I_{X_+}$ $I_{X_+}$  and $I_{XX}$ $/$ $I_{X_-}$ $I_{X_-}$ (Supplementary Table I), which is consistent with our assignment that IXX2 is a $X_+X_-$ biexciton.

Although expected, single photon emission was not demonstrated here due to weak emission. Future experiments incorporating localized interlayer excitons in plasmonic nanocavities should be able to enhance the emission rate, as was recently demonstrated for monolayer quantum emitters~\cite{LuoNNano2018}. Thus, localized interlayer excitons with finite dipole moment constitute quantum emitters with field tuneable energy over a wide range. Moreover, their biexciton cascade emission is devoid of the fine structure splitting which limits the entanglement fidelity of the emitted photon pair in intralayer excitons. In addition, due to their static dipole moment, electrostatic confinement seems to be the natural way to realize an array of quantum emitters with the on-site dipole repulsion leading to the Mott-phase of the Bose-Hubbard model~\cite{YuSciAdv2017,WuPRL2018}. Another interesting possibility is to confine many dipolar excitons in a larger trap (tens of nm). With increasing exciton density a crystallization phase transition into 1D or 2D dipolar crystal might occur, which could be probed through their phonon modes~\cite{RablPRA2007}. A dipolar crystal of an interlayer excitonic condensate might give rise to more exotic many-body states such as supersolids while the internal spin-valley degree of freedom could lead to magnetic instabilities and frustration. \\

\justify

\textbf{Acknowledgments} We acknowledge many enlightening discussions with Ata\c{c} Imamo\u{g}lu and Martin Kroner. We also thank Robert Lemasters and Hayk Harutyunyan for help on atomic layer deposition.  A. S. acknowledges support from NSF through the EFRI program-grant \# EFMA-1741691. 

\justify
\textbf{Author Contributions}  ~$^\dagger$ W. L. and X. L. contributed equally to this work.  A. S., W. L., X. L. and S. D. conceived the project. W. L., X. L., S. D. and L. D.carried out the measurements. W. L. performed the theoretical calculations. X. L., S. D. and L. D. prepared the samples. A. S. supervised the project. All authors were involved in analysis of the experimental data and contributed extensively to this work.\\

\textbf{Methods}
\\
\textbf{Sample fabrication} We use electron beam lithography and thermal evaporator to fabricate the pillar arrays (5~nmCr/85~nmAu) or electrical contacts (5~nmCr/55~nmAu) on 300~nm SiO$_2$/Si substrate. A thin layer of SiO$_2$ (3~nm) is subsequently deposited on pillars by using atomic layer deposition as the spacer. We transfer the mechanically exfoliated samples by polydimethylsiloxane-based dry transfer method on the as-patterned pillars/electrical contacts, with monolayer WSe$_2$ (HQ graphene) on top of monolayer MoSe$_2$ (HQ graphene). The sample with electrical contacts is encapsulated between two hexagonal boron nitride (HQ graphene) layers. Graphite (NGS) layers are used as the bottom gate and top gates. After the stacking of top MoSe$_2$, the sample is annealed in 5$\%$ H$_2$/95$\%$ N$_2$ at 125$^{\circ}$C for 2~h. 


\justify
\textbf{Photoluminescence spectroscopy} Photoluminescence spectroscopy is based on two home-built low temperature  microscope set-ups. 
The WSe$_2$/MoSe$_2$ sample is loaded into a closed-cycle cryostat (AttoDry 800) with the electrical connection and then into another cryostat (BlueFors cryogenics) with magnetic field from -8 T to +8 T. Both of them are cooled to $\sim$4 K all conducted measurements.
A piezoelectric controller (Attocube systems) is used to position the sample.
The emission is collected using an achromatic lens (NA = 0.42 for AttoDry 800 and NA = 0.63 for BlueFors cryogenics) and directed to a high-resolution (focal length: 500 mm for AttoDry 800 and 750 mm for BlueFors cryogenics) spectrometer (Princeton Instrument HR-500 for AttoDry 800 and Princeton Instruments SP-2750 for BlueFors cryogenics) where it is dispersed by a 1200 g/mm or 300 g/m grating (both blazed at 750 nm). 
A charge coupled device (Princeton Instrument PIXIS-400 CCD for AttoDry 800 and PyLoN CCD for BlueFors cryogenics) is used as a detector.
The excitation laser is a mode-hop-free tunable continuous-wave Ti:Sapphire laser (MSquared Lasers) with resolution of 0.1 pm are coupled to a single-mode fiber to clean the mode.
The Ti:Sapphire laser has spot size of $\sim$ 1$\mu$m.
The polarization of the incident laser is controlled using a polarizer together with a liquid crystal variable retarder (LCR).
The polarization detection is performed by using a $\lambda/4$ waveplate (achromatic, 690-1200 nm) placed before the Wollaston prism. 
The circularly polarized emission is converted into linearly polarized light through the $\lambda/4$ waveplate.
The s- and p- components of linear polarized light is then displaced by the Wollaston prism. Another achromatic $\lambda/4$ waveplate is placed after the Wollaston prism to convert the linearly polarized light into a circularly polarized signal, in order to avoid the sensitivity to the grating efficiency. In all measurements, the magnetic field $\it{B}$ is applied perpendicular to the plane of the sample and the voltage is applied to the graphite gates with a Keithley 2400 sourcemeter.

\clearpage

\textbf{\underline{Supplementary Information Contents:}}\\
Figure S1. Time-trace photoluminescence (PL) emission of localized interlayer excitons.\\
Figure S2. Linewidths from from localized interlayer excitons.\\
Figure S3. Electrical tuning of IX3 group.\\
Figure S4. PL of IX3 group under different excitation powers.\\
Figure S5. Localized interlayer exciton IX4 and biexction IXX4.\\
Figure S6. Power dependence of localized interlayer excitons and biexcitons.\\
Figure S7. Multi-excitonic complex of localized interlayer excitons.\\
Figure S8. PL of localized interlayer exciton IX5 group under different excitation powers.\\
Figure S9. {\it g}-factors of localized interlayer excitons in IX5 group.\\
Table I. Correlation of biexciton and exciton intensities for IX2-IXX2.\\
Note 1. Estimation of confinement length for exciton IX.\\
Note 2. Estimation of exchange energy for biexciton IXX.\\
Note 3. Estimation of total energy of triexciton IXXX.\\


\clearpage
\justify

\begin{figure}[H]
\centering
\includegraphics[width=90mm]{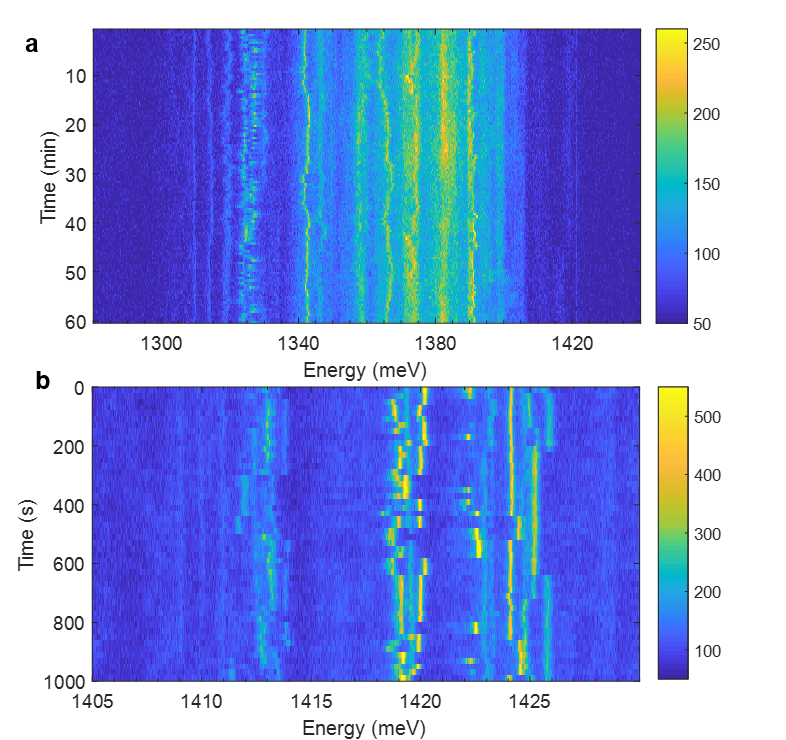}\\
\justify
{\bf Figure S1: Time-trace PL emission of localized interlayer excitons in MoSe$_2$/WSe$_2$ heterostructure.} Sharp peaks spectrally wander in the range of 1300 to 1430~meV, indicating that they are originating from interlayer excitons. Excitation is linearly-polarized, with wavelength $\lambda$ = 740~nm in panel {\bf a}, $\lambda$ = 735~nm in panel {\bf b}. Incident power $P$ = 30~nW in panel {\bf a}, $P$ = 20~nW in panel {\bf b}.
\end{figure}

\begin{figure}[H]
\centering
\includegraphics[width=100mm]{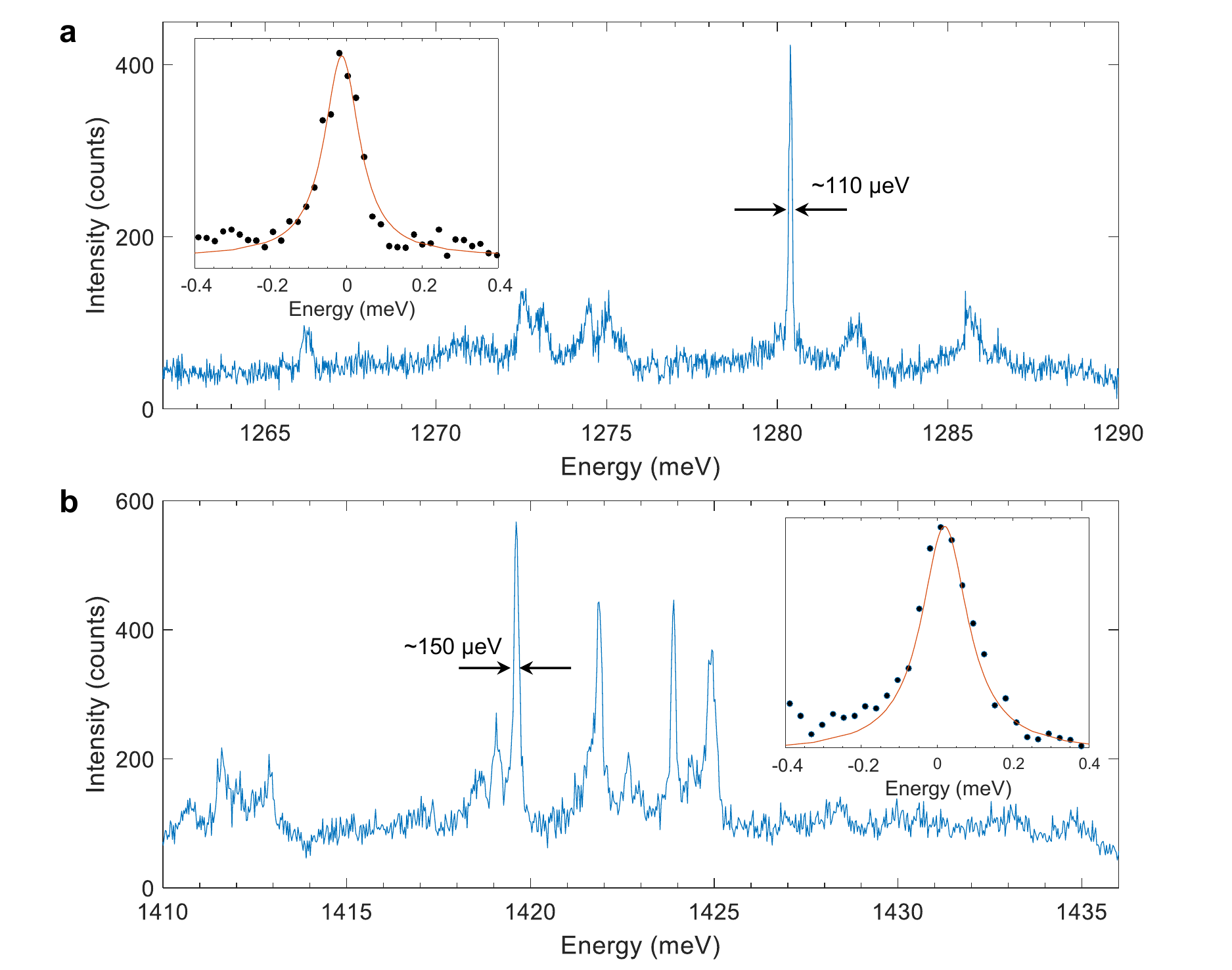}\\
\justify
{\bf Figure S2: Linewidths from localized interlayer excitons.} PL spectra of localized interlayer excitons showing sharp linewidths of $\sim$110~$\mu$eV (a) and $\sim$150~$\mu$eV (b). Inset: fitting with Lorentzian function for localized interlayer excitons at $\sim$1280~meV (a) and $\sim$1420~meV (b). Excitation is linearly-polarized, with wavelength $\lambda$ = 770~nm in panel {\bf a}, $\lambda$ = 735~nm in panel {\bf b}. Incident power $P$ = 6.5~nW in panel {\bf a}, $P$ = 20~nW in panel {\bf b}.
\end{figure}

\begin{figure}[H]
\centering
\includegraphics[width=200mm]{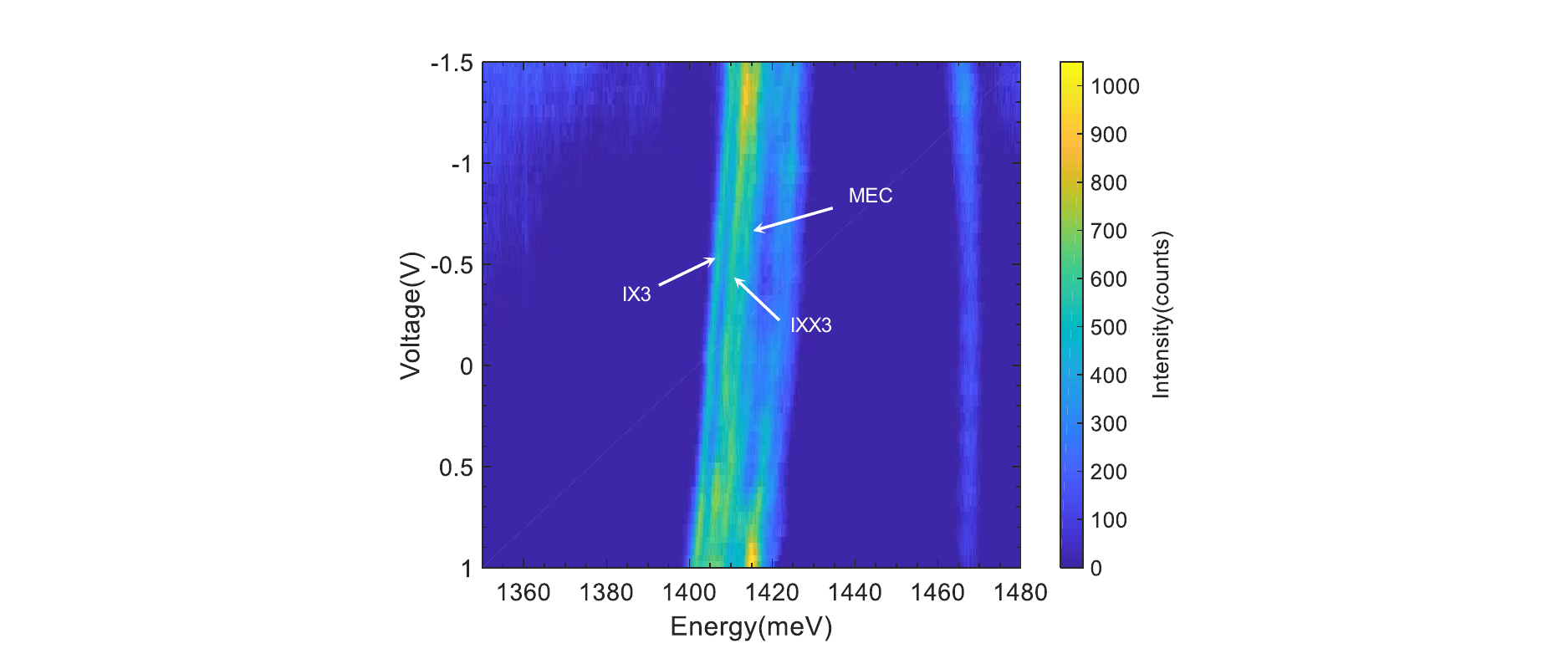}\\
\justify
{\bf Figure S3: Electrical tuning of IX3 group.} Exciton IX3, biexciton IXX3 and multi-excitonic complex (MEC) show the same tuning rate as bottom gate voltage changes from -1.5~V to +1~V. Excitation is linearly-polarized, with wavelength $\lambda$ = 758~nm. Incident power $P$ = 8~$\mu$W.
\end{figure}

\begin{figure}[H]
\centering
\includegraphics[width=160mm]{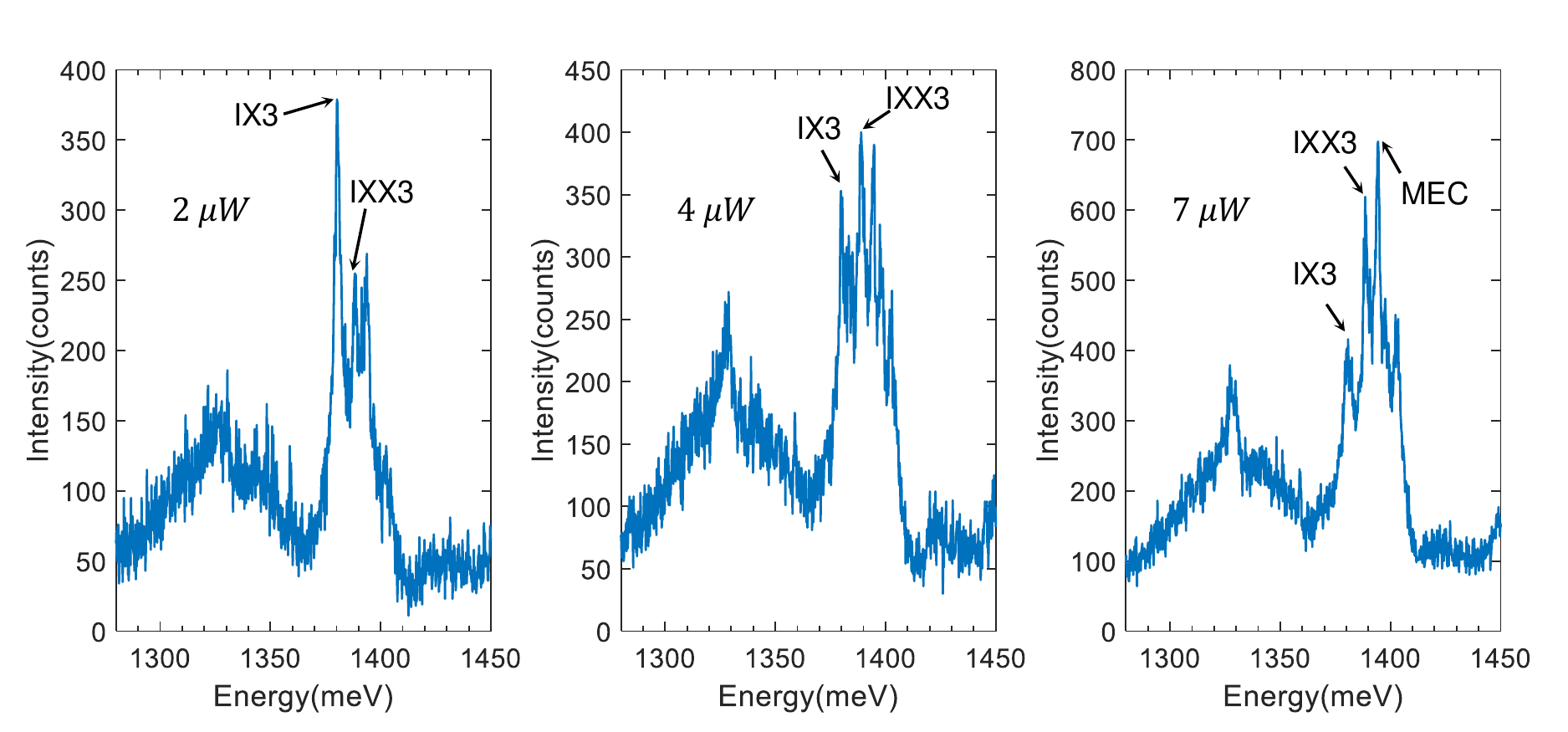}\\
\justify
{\bf Figure S4: PL spectra of IX3 group under different excitation powers.} At low power $P$ = 2~$\mu$W, exciton IX3 is stronger than biexciton IXX3 (left). IXX3 is stronger than IX3 when $P$ = 4~$\mu$W (middle). MEC appears and dominate when $P$ = 7~$\mu$W (right). Excitation is linearly-polarized, with wavelength $\lambda$ = 765~nm.
\end{figure}

\begin{figure}[H]
\centering
\includegraphics[width=120mm]{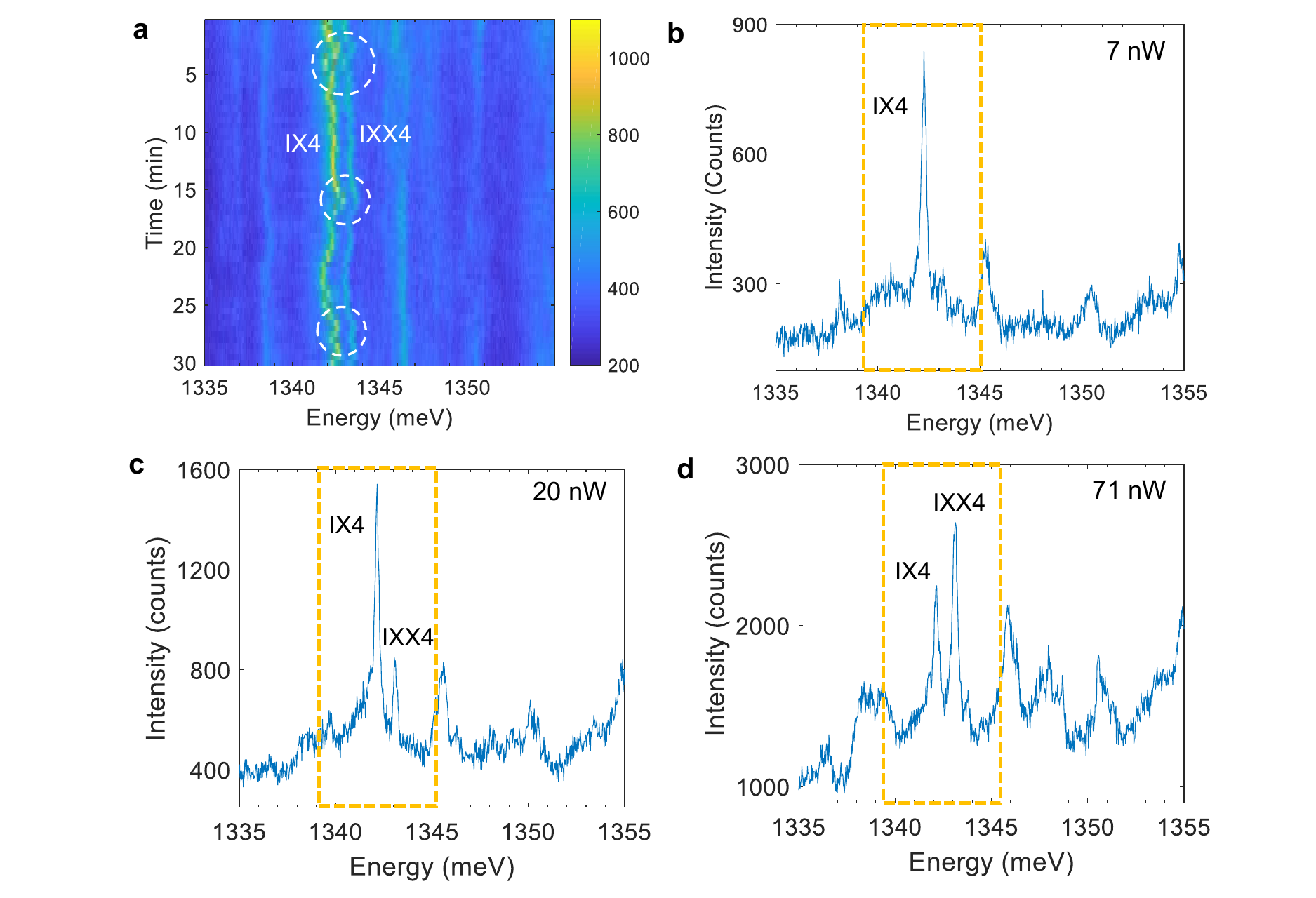}\\
\justify
{\bf Figure S5: Localized interlayer exciton IX4 and biexction IXX4.} {\bf a,} Time-trace PL emission of interlayer exciton IX4 and biexciton IXX4. Dashed circles highlight the same spectral jittering patterns, which implies that IX4 and IXX4 are correlated. {\bf b-d} PL spectra of IX4-IXX4 under different excitation powers, 7~nW (b), 20~nW (c) and 71~nW (d). Only exciton IX4 appears at low power (7~nW). Biexciton IXX4 shows up at intermediate power (20~nW) and becomes stronger than IX4 at high power (71~nW). Excitation laser is linearly-polarized, with wavelength $\lambda$ = 745~nm. Incident power $P$ = 50~nW in panel {\bf a}.
\end{figure}

\begin{figure}[H]
\centering
\includegraphics[width=120mm]{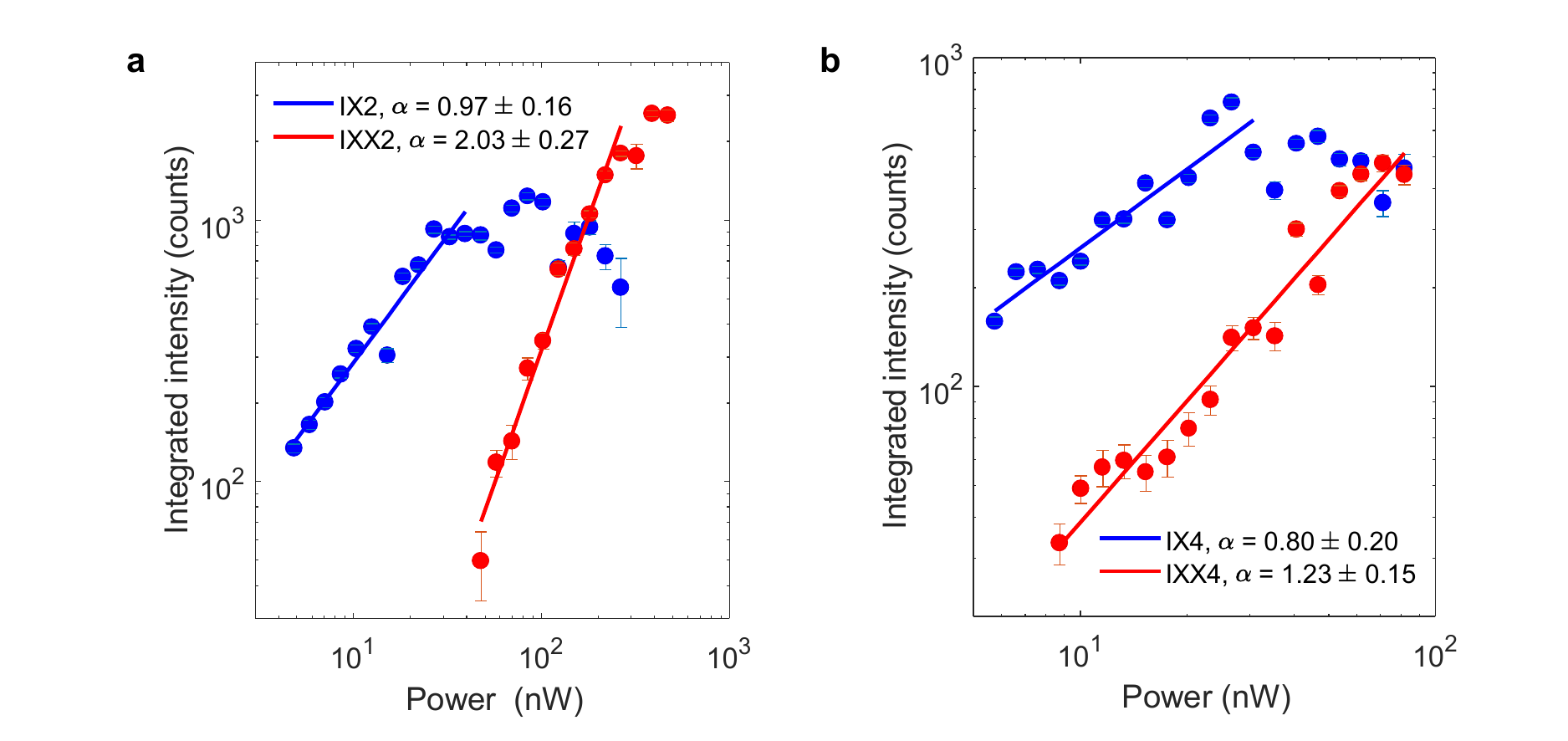}\\
\justify
{\bf Figure S6: Power-dependence of localized interlayer excitons and biexcitons.} Integrated intensity of emission peaks in IX2-IXX2 (a) and IX4-IXX4 (b). The fitting was done with a power law function, $I$ $\propto$ $P^\alpha$. We obtain $\alpha (IXX) ≈ \approx 2\alpha (IX)$, indicating that IXX2 and IXX4 are biexcitons. 
\end{figure}

\begin{figure}[H]
\centering
\includegraphics[width=120mm]{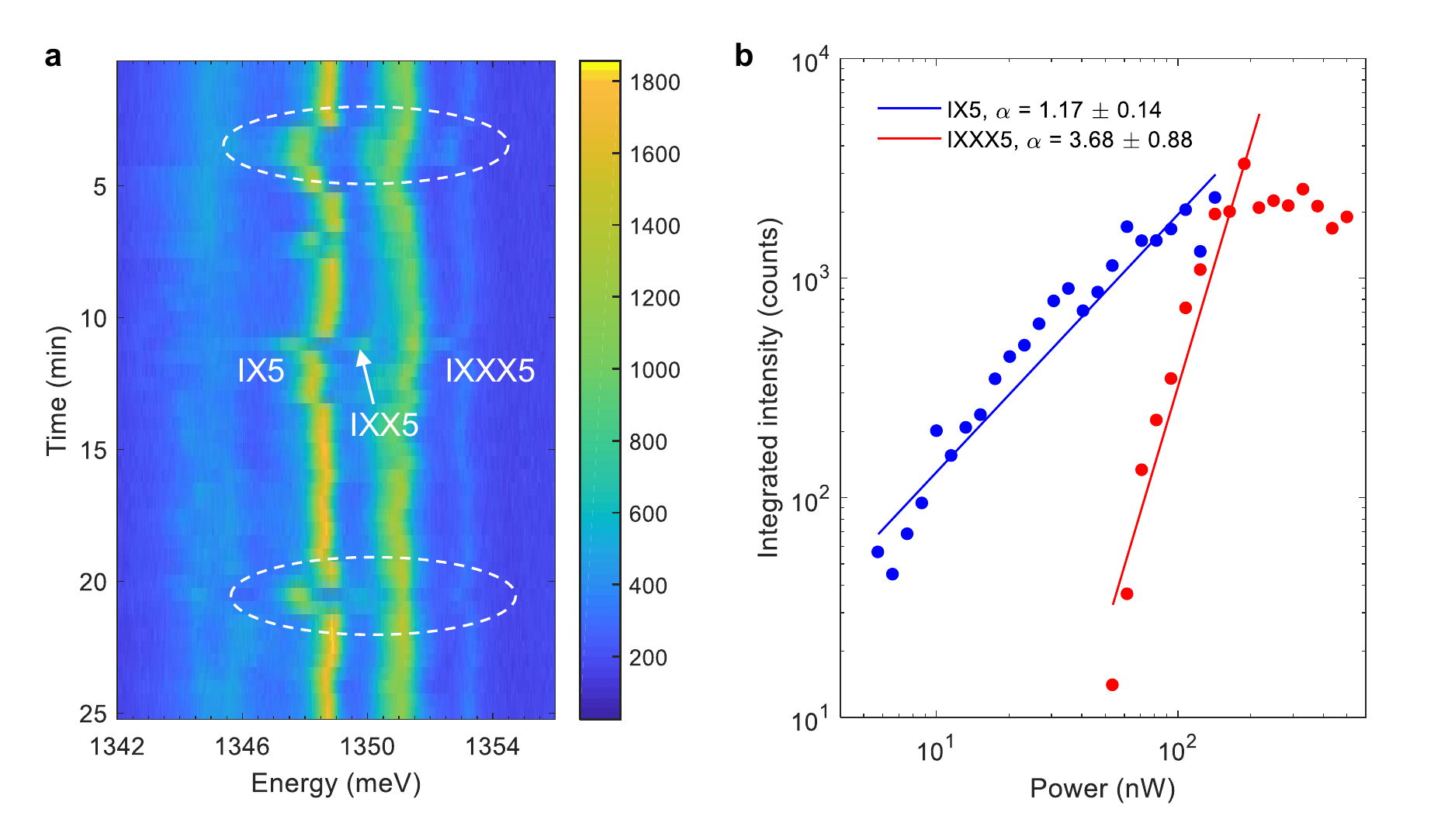}\\
\justify
{\bf Figure S7: Multi-excitonic complex of localized interlayer excitons.}  {\bf a,} Time-trace PL emission of interlayer exciton IX5, biexciton IXX5 and triexciton IXXX5. Dashed circles highlight the same spectral jittering patterns. Energy spacings between IX5 $\&$ IXX5 and IX5  $\&$ IXXX5 are $\sim$2.3~meV and $\sim$4.7~meV, respectively.  {\bf b,} Integrated intensity of emission peaks IX5 and IXXX5.  We obtain $\alpha (IXXX5) ≈ \approx 3\alpha (IX)$ from power-law fitting, indicating that IXXX5 is a triexciton. As the emission energy of IXX5 is very close to another uncorrelated and strong peak (shown in a), we can not extract the power-dependent integrated intensity for IXX5. Excitation is linearly-polarized, with wavelength $\lambda$ = 725~nm in panel {\bf a},  $\lambda$ = 745~nm in panel {\bf b}. Incident power $P$ = 45~nW in panel {\bf a}.
\end{figure}

\begin{figure}[H]
\centering
\includegraphics[width=130mm]{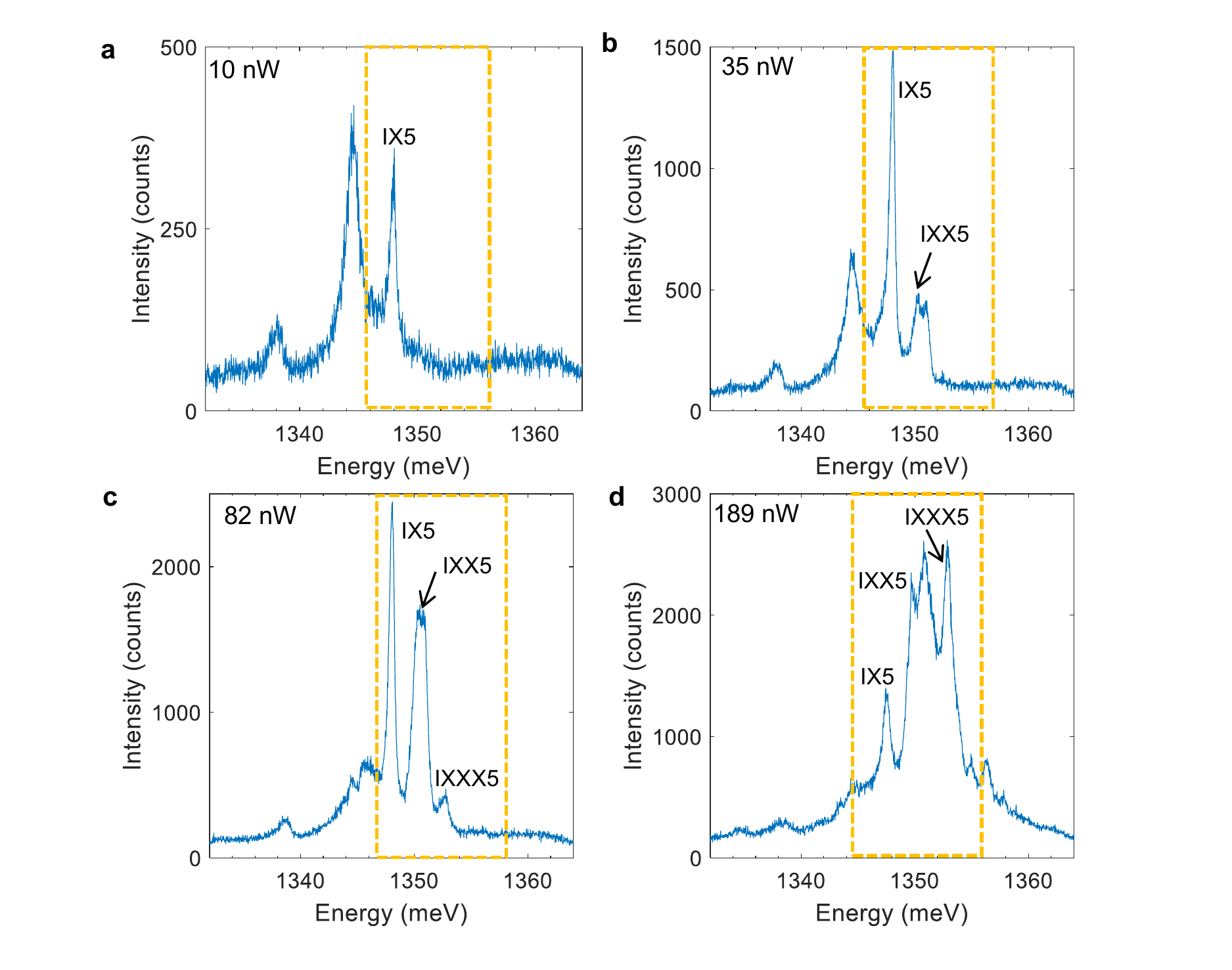}\\
\justify
{\bf Figure S8: PL emission of interlayer exciton IX5 group under difference excitation powers.}  Only exciton IX5 shows up at low power, 10~nW(a). Biexciton IX5 starts to appear at $P$ = 35~nW (b), and triexciton IXXX5 is activated at higher power 82~nW (c). At $P$ = 189~nW (d), triexciton IXXX5 is stronger than biexciton IXX5, and exciton IX5 is the weakest. Excitation is linearly-polarized, with wavelength $\lambda$ = 745~nm.
\end{figure}

\begin{figure}[H]
\centering
\includegraphics[width=140mm]{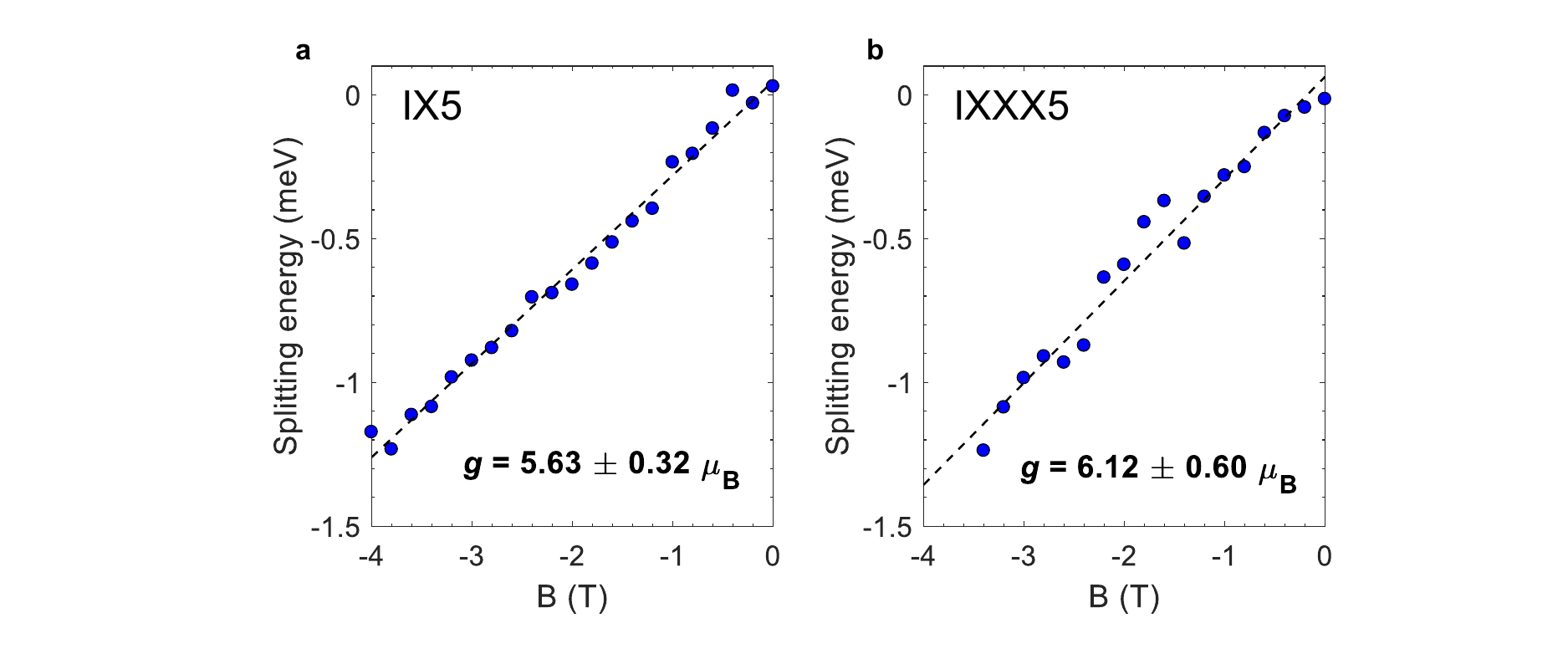}\\
\justify
{\bf Figure S9: $g$-factors of localized interlayer excitons in IX5 group.}  Exciton IX5 (a) and triexciton IXXX5 (b) exhibit the same value of $g$-factor. Excitation is linearly-polarized, with wavelength $\lambda$ = 745~nm. Incident power $P$ = 180~nW.
\end{figure}

\begin{table}[H]
\caption{Correlation of biexciton and excition intensities for IX2-IXX2}
\centering 
\begin{tabular}{c c c c}
\hline\hline 
& $\sigma^-$ excitation & $\sigma^+$ excitation &  Difference$^2$ \\ 
\hline
DCP of exciton$^1$  & -0.19& -0.29 &   -0.10\\
$I_{XX}$ $/$ $I_{X_+}$$I_{X_-}$    & 11.035e-05 & 10.823e-05  & -0.48$\%$  \\
$I_{XX}$ $/$ $I_{X_+}$$I_{X_+}$   & 1.6543e-04 & 2.0172e-04 &   4.94$\%$ \\
$I_{XX}$ $/$ $I_{X_-}$$I_{X_-}$  & 7.4878e-05 & 5.9556e-05 &   -5.70$\%$  \\
\hline
\end{tabular}
\end{table}

$^1$ Degree of circular polarization (DCP) of exciton, is obtained by calculating ($I_{\sigma^+}$ - $I_{\sigma^-}$) / ($I_{\sigma^+}$ + $I_{\sigma^-}$).\\
$^2$ Difference of DCP is calculated by:  DCP($\sigma^+$) - DCP($\sigma^-$). Differences of Ratio $R$, such as $R$ of $I_{XX}$ $/$ $I_{X_+}$$I_{X_-}$,  is obtained by calculating 2($R$($\sigma^+$) - $R$($\sigma^-$)) / ($R$($\sigma^+$) + $R$($\sigma^-$)).\\

\textbf{Note 1: Estimation of the confinement length forexciton  IX \cite{SchinnerPRL2013}}
\\
According to the dipolar-crystal-like model, the overlap of the excitonic wavefunctions is neglected, 
so the total exciton energy in a parabolic trap is given by
$$E_{NIX}=N E_{IX}^0+\sum_i \frac{M\Omega^2}{2}r_i^2+\sum_{i<j,i,j=1,...,N}\frac{d^2}{\epsilon_r r_{i,j}^3},$$ where $N$ is the number of excitons, $E_{IX}^0$ is the optical energy of the exciton,
$\Omega$ is the characteristic confinement frequency of an exciton in the trap.
$M = m_e^*+m_h^*$ is the total exciton mass, d is the interlayer spacing, $r_i$ is the coordinate of i-exciton, and $r_{i,j}$ is the distance between i-, j-exciton. 
In order to calculate the emission energy of biexciton, we only need to consider the single exciton and biexciton energy
$$E_{IX}= E_{IX}^0+\frac{M\Omega^2}{2}r_1^2,$$
$$E_{IXX}= 2E_{IX}^0+\frac{M\Omega^2r_1^2}{2}+\frac{M\Omega^2r_2^2}{2}+\frac{d^2}{\epsilon_r r_{1,2}^3}.$$ The emission energy of biexciton is given by
$$\hbar \omega_{IXX}= E_{IXX,\mathrm{min}}-E_{IX,\mathrm{min}},$$
where the $E_{IX,\mathrm{min}}$, $E_{IXX,\mathrm{min}}$ are the minimum of $E_{IX}$, $E_{IXX}$, that is, the ground states of IX and IXX respectively.
For the exciton IX, the ground state is that the exciton stays at the lowest potential energy; for the biexciton IXX, due to the repulsive onsite dipole-dipole interaction ($U_\mathrm{dd}^\mathrm{on-site}$) between two excitons, the ground state is a diatomic geometry with $r_1=-r_2=\frac{1}{2}r_{1,2}=R_{IXX}$,
$$E_{IX,\mathrm{min}}= E_{IX}^0,$$
$$E_{IXX,\mathrm{min}}= 2E_{IX}^0+2\frac{M\Omega^2R_{eq,IXX}^2}{2}+\frac{d^2}{8\epsilon_r R_{eq,IXX}^3},$$
where $R_{eq,IXX}=(\frac{3d^2}{16\epsilon_r M\Omega^2})^{\frac{1}{5}}$ is the solution for $\frac{dE_{IXX}}{dR_{IXX}}=0$.
Then we can get $$\hbar \omega_{IXX}= E_{IX}^0+[(\frac{3}{16})^{\frac{2}{5}}+(\frac{1}{6})^{\frac{3}{5}}](\frac{M^3\Omega^6 d^4}{\epsilon_r^2})^{\frac{1}{5}}.$$
Therefore, the energy spacing between the IX and IXX emission is 
$\Delta E=[(\frac{3}{16})^{\frac{2}{5}}+(\frac{1}{6})^{\frac{3}{5}}](\frac{M^3\Omega^6 d^4}{\epsilon_r^2})^{\frac{1}{5}}.$

In the WSe$_2$/MoSe$_2$ heterobilayer, $M=m_e^*+m_h^*\approx1.15m_0$, where $m_0$ is the electron bare mass \cite{LarentisPRB2018,WuPRL2018}, $d = 0.7$~nm \cite{RiveraScience2016}, $\epsilon_r\approx(1+3.9)/2=2.45$ \cite{ChernikovPRL2014} for the SiO$_2$ substrate. Considering the energy spacings $\Delta E$ in Fig. 3a $\&$ 3b from main text are $\sim$ 2~meV, we obtain the parabolic confinement frequency $\Omega \sim 2.51$~meV. The confinement frequency $\Omega$ is related to the confinement length $l$ by
$\Omega=\frac{\hbar}{M l^2}$, so the confinement length for our interlayer quantum emitter is $\sim$ 5.14~nm. The Bohr radius $a_B$ for IX is $\sim 2$~nm~\cite{YuSciAdv2017}, smaller than the confinement length, which validates the model.\\

\textbf{Note 2: Estimation of the exchange energy of biexciton IXX~\cite{YuSciAdv2017,RiveraScience2016}}
\\
The on-site Coulomb interaction between two IX wavepackets in the same quantum emitter can be separated into two parts: 
(i) dipole-dipole interaction  $U_\mathrm{dd}^\mathrm{on-site}$ regardless of the valley index ($|X_+X_-\rangle$, $|X_+X_+\rangle$ and $|X_-X_-\rangle$),
 $$U_\mathrm{dd}^\mathrm{on-site} = \frac{1}{2\pi}(\frac{a_B}{w})^2 \frac{d}{a_B} E_b,$$
(ii) Exchange interaction $U_{ex}$ between excitons in the same valley only ($|X_+X_+\rangle$ and $|X_-X_-\rangle$), 
$$U_{ex} = \frac{1}{2\pi}(\frac{a_B}{w})^2 E_b,$$
where $a_B$ is the exciton Bohr radius, $w$ is the real space extension of the exciton center of mass wavefunction, $E_b$ is the interlayer exciton binding energy. The energy spacing between IXX and IX in the main text is actually $U_\mathrm{dd}^\mathrm{on-site}$ of 2.0~meV, since no exchange interaction exists in $|X_+X_-\rangle$.
Using $E_b = 0.2$~eV, $a_B=2$~nm, we have $w\sim$ 4.7~nm, $U_{ex} \sim 5.7$~meV. Large $U_{ex}$ makes the $|X_+X_+\rangle$ or $|X_-X_-\rangle$ have much higher energy than $|X_+X_-\rangle$ and not favorable due to thermalization. \\

\textbf{Note 3: Estimation of total energy of triexciton IXXX \cite{SchinnerPRL2013}}
\\
Let us use the above model to consider the transition from $|X_+X_+X_-\rangle$ (or $|X_-X_-X_+\rangle$) to $|X_+X_-\rangle$ and then from $|X_+X_-\rangle$ to either $|X_+\rangle$ or $|X_-\rangle.$
The exciton enegy is $$E_{IX,min}= E_{IX}^0,$$
The biexciton energy is $$E_{IXX,min}= 2E_{IX}^0+U_{dd,2}=2E_{IX}^0+2.3meV, $$ 
and the triexciton energy is $$E_{IXXX,min}= 3E_{IX}^0+3 U_{dd,3} +U_{ex}= 3E_{IX}^0+3\frac{M\Omega^2r_{eq,3}^2}{2}+\frac{\sqrt{3}e^2d^2}{3\epsilon r_{eq,3}^3}+U_{ex}$$
$$=3E_{IX}^0+5.4 meV+U_{ex}, r_{eq,3}=(\frac{\sqrt{3}e^2d^2}{3\epsilon M\Omega^2})^{\frac{1}{5}},$$
where $U_{dd,2}$ is the dipole-dipole interaction at the biexcton states IXX with a diatomic geometry, which is from the experiment and gives the value of $\Omega$. $U_{dd,3}$ is the calculated dipole-dipole interaction between each two exciton at the triexcton states IXXX with a symmetric triangular geometry, and $U_{ex}$ is the exchange energy between the two dipoles in the same valley. Here we ignore the reorganization of the IXXX by the exchange energy.

If $U_{ex}=0$, we can get the lower bound of $E_{IXXX,min}$ to be $3E_{IX}^0+5.4$ meV; if we use the same relation between $U_{ex}$ and $U_{dd}$ as in the biexciton states, which is $U_{ex}=\frac{a_B}{d}U_{dd,3}=\frac{2}{0.7}\frac{5.4}{3}=5.14$ meV with $a_B = 2$ nm and $d = 0.7$ nm, the higher bound of $E_{IXXX,min}$ is estimated to be $3E_{IX}^0+10.54$ meV.

In the experiment spectra with IX5, IXX5, and IXXX5, 
$$E_{IXX,min}-E_{IX,min}=\hbar \omega_{IXX}=E_{IX}^0+2.3 meV,$$
$$E_{IXXX,min}-E_{IXX,min}=\hbar \omega_{IXXX}=E_{IX}^0+4.7 meV,$$
Adding the two equations, we have the experiment result is $E_{IXXX,min}=3E_{IX}^0+7.0$~meV, within the estimated range.

If we consider the reorganization, IXXX becomes an asymmetric configuration to lower the total energy of IXXX, which could match the experiment result. In other words, the estimation of $U_{ex}=\frac{a_B}{d}U_{dd,3}$ is resonable.\\


\end{document}